\documentclass[journal,onecolumn]{IEEEtran}
\IEEEoverridecommandlockouts
\usepackage{cite}
\usepackage{amsmath,amssymb,amsfonts}
\usepackage{algorithmic}
\usepackage{graphicx}
\usepackage{hyperref}
\usepackage{cleveref}
\usepackage{epstopdf}
\usepackage{mathtools}
\usepackage[acronym,toc,shortcuts]{glossaries}
\usepackage{caption}
\usepackage{authblk}
\usepackage{algorithm}
\renewcommand{\baselinestretch}{1.19}
\usepackage{subfigure}

\newacronym{WBAN}{WBAN}{wireless body area network}
\newacronym{SN}{SN}{sensor node}
\newacronym{IC}{IC}{integrated circuit}
\newacronym{QoS}{QoS}{quality-of-service}
\newacronym{LP}{LP}{linear program}
\newacronym{LFP}{LFP}{Linear Fractional Program}
\newacronym{EE}{EE}{energy efficiency}
\def\BibTeX{{\rm B\kern-.05em{\sc i\kern-.025em b}\kern-.08em
    T\kern-.1667em\lower.7ex\hbox{E}\kern-.125emX}}
\begin{document}

\title{Energy Efficiency Maximization of Self-Sustained Wireless Body Area Sensor Networks\\
}

\author{Osama~Amjad, Ebrahim~Bedeer, \textit{Member, IEEE}, and, Salama~Ikki, \textit{Senior Member, IEEE}}



\maketitle

\begin{abstract}
\noindent
\textbf{ Electronic health monitoring is one of the major applications of wireless body area networks (WBANs) that helps with early detection of any abnormal physiological symptoms. In this paper, we propose and solve an optimization problem that maximizes the energy efficiency (EE) of WBAN consisting of sensor nodes (SNs) equipped with energy harvesting capabilities communicating with an aggregator. We exploit the structure of the optimization problem to provide a sub-optimal solution at a lower computational complexity and derive the mathematical expressions of upper and lower bounds of the source rates of the SN. The simulation results reveal that the optimal allocation of the source rate to energy critical SNs improves the system performance of WBAN in terms of energy efficiency during different everyday activities.}
\end{abstract}
\section{Introduction}
Electronic Health (eHealth) monitoring systems with wireless body area networks (WBANs) help integrate the patient's data processing and communications technologies into traditional medical facilities and serves as a promising approach to boost the health-care efficiency. In WBAN, the sensor nodes (SNs) monitor the patient's vital signs and send the data wirelessly to the aggregator \cite{movassaghi2014wireless}. These SNs are conventionally powered by batteries, which are needed to be replaced once the energy is consumed. Therefore, wireless energy harvesting serves as an alternative approach that enables self-sustained SNs operations by scavenging energy from biomechanical, biochemical, and ambient sources (e.g., thermal, electromagnetic radiations) \cite{akhtar2017energy}.

Due to limited battery life, saving energy of the SNs is of significant importance. Therefore, WBAN has to provide sustainable battery lifetime, high energy efficiency (EE), and quality-of-service (QoS) of the data stream. In \cite{ibarra2016qos}, an efficient power QoS control scheme for energy harvesting WBAN is proposed that ensures the best possible QoS by efficiently transmitting the data packets. Stochastic modeling of wirelessly powered wearables proposed in \cite{mekikis2017stochastic} provides an analytical framework for the SNs ability to notify the medical staff about the patient's condition promptly. In \cite{yang2018wireless}, a medium access control (MAC) layer protocol for WBAN is proposed that utilizes the CSMA/CA-TDMA hybrid schemes to extend the lifetime and EE of SNs by saving energy. In \cite{liu2017medium}, a MAC protocol for WBAN is proposed that ensures QoS and EE in the power constrained network by dynamically optimizing the transmission slot such as the energy consumption of the SNs is minimized. Cooperative energy harvesting-adaptive MAC protocol proposed in \cite{esteves2015cooperative} improves the WBAN performance in terms of delay, EE, and throughput by changing its operation based on the energy harvesting conditions.

Compared to the existing work, this paper aims to maximize the overall EE of the energy harvesting WBAN. In particular, we formulate and solve a novel optimization problem that optimally allocates each SN source rate to maximize the EE subject to power budget and limited energy harvesting constraints. The linear fractional EE problem is converted to an equivalent linear form by using the Charnes-Cooper transformation. Moreover, we exploit the structure of the EE optimization problem to provide a sub-optimal solution at lower computational complexity and derive the upper and lower bounds of the source rates. Finally, we evaluate the performance of the proposed schemes through extensive simulations to validate the proficiency and performance merits in terms of EE of the system.
\section{System Model}
We consider a WBAN with one aggregator carried by the patient which acts as the gateway. There are $N$ SNs in the network deployed on the patient's body in a star topology such that each node directly communicates with the aggregator. The aggregator is assumed to be connected with a reliable energy source, whereas the SNs are energy critical and are supposed to have a rechargeable battery that has an ability to harvest energy from biochemical and biomechanical energy sources available in the human body.

In WBAN, each SN inquires a dedicated guaranteed time slot from the aggregator, during which it periodically transmits its data using standard IEEE 802.15.4 TDMA scheme \cite{yang2018wireless}. In each time slot, based on the state of the patient during different activities (e.g., relaxing, walking, running, etc.) the SN $i$ can harvest different amount of energy in a range of [$E_i^{\min}, E_i^{\max}$], denoting the minimum energy level required to be maintained and the maximum battery capacity, respectively.

Since the source rate and the energy recharging rate during each time slot remains constant, the energy harvesting process at SN $i$ can be modeled as a discrete-time Markov chain \cite{ibarra2016qos}, \cite{he2011optimal} represented as \{$\mathbf{A}_i, \mathbf{P}_i$\}, where $\mathbf{A}_i$ is the set of states in the Markov chain model, and $\mathbf{P}_i$ is the transition probability matrix. The energy recharging rate at state $m$ ($m\in$ $\mathbf{A}_i$) is expressed as $g_{i}^{(m)}$. Fig. 1 shows the two state Markov chain with transition probabilities $P_{10}$, and $P_{01}$ from state $S_1$ to $S_2$, and from state $S_2$ to $S_1$, respectively. The states in $\mathbf{A}_i$ are arranged in an increasing order as $g_{i}^{(1)}\leq g_{i}^{(2)}\dots \leq g_{i}^{\mathbf {|A}_i|}$, where $\mathbf{|A}_i|$ is the cardinality of $\mathbf{A}_i$ and represents the number of the states in $\mathbf{A}_i$.  Let $\mathbf \Pi_{i}$ be the steady-probability vector at SN $i$, that can be calculated as follows: $\mathbf \Pi_{i}^{T}\mathbf{P}_i$ = $\mathbf \Pi_{i}^{T}$, and $\mathbf \Pi_{i}^{T}\mathbf I = 1$, where $\mathbf I$ is the identity vector with all entries equal to 1. The long-term average recharging rate of SN $i$ is then written as $g_{i}^{avg}$ = $\mathbf \Pi_{i}^{T} \mathbf g_{i}$, where $\mathbf g_{i}$ is the vector of the recharging rates at sensor $i$.

\begin{figure}[t]
\centering
\captionsetup{justification=centering}
\includegraphics[width=0.4\textwidth]{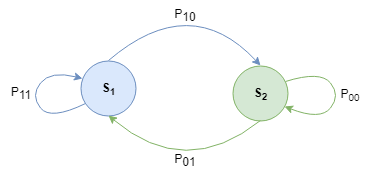}
\caption{\label{fig:data}Two state discrete Markov chain of energy harvesting process.}
\end{figure}

In WBAN, the power consumption of a SN depends on: sensing power consumption and transmission power consumption. The SN consumes energy while capturing the readings from the human body, and eventually sending these wirelessly as data packets to the aggregator. Therefore, the sensing power consumption at a SN $i$ is proportional to the source rate $r_i$ \cite{he2011optimal} denoted by $P_{s, i}= \psi_i r_i$, where $\psi_i$ is the energy cost of sensing. In the eHealth system, random body movements of the patient cause changes in the distance and the direction of the SN to the aggregator that results in the change of path loss and ultimately attenuates the transmission power consumption $P_{t, i}$. Therefore, according to the radio energy model in \cite{heinzelman2000energy}, to guarantee a certain minimum received power at the aggregator, the $P_{t, i}$ takes $d_i^{m_p}$ as a path loss and the energy cost due to a channel variation in respect with distance $d_i$ between SN $i$ and the aggregator, and thus makes the received power independent of $d_i^{m_p}$. The transmission power consumption depends on the path loss model of the wireless channels in WBAN as illustrated in \cite{heinzelman2000energy}
\begin{equation}
\small\begin{aligned}
\hspace{-6pt}PL(d_i)=PL(d_o)+10 \hspace{2pt}m_p \log_{10}\Big(\frac{d_i}{d_o}\Big)+X_\sigma, \hspace{7pt}\forall i \in N,
\end{aligned}\normalsize
\label{equ1}
\end{equation}
where $PL(d_o)$ is the path loss at reference distance $d_o$, $m_p$ is the path loss exponent. The transmission power consumption at SN $i$ can be modeled as \small$P_{t, i}=X_\sigma \beta_i r_i+\theta_i$\normalsize, where $\theta_i$ is the constant energy cost of transmit electronics of sensor $i$, $\beta_i$ is the transmission energy consumption cost of SN $i$ given by \small$\beta_i=\zeta_i d_i^{m_p}$\normalsize and, $\zeta_i$ is a coefficient term associated with the energy cost of transmit amplifier. $X_\sigma$ is the Gaussian random variable that represents shadowing, and denoted as $\mathcal{N}(0, \sigma_s^2)$. The standard deviation $\sigma_s$ depicts the different postures of the body such as relaxing, walking or running \cite{d2010statistical}. The total power consumption at sensor $i$ is the sum of $P_{s, i}$ and $P_{t, i}$ given as
\begin{equation}
\small\begin{aligned}
P_i&=P_{s, i}+P_{t, i}=\psi_i r_i+X_\sigma\beta_i r_i + \theta_i,\\
 &=(\psi_i+X_\sigma \zeta_i d_i^{m_p})r_i+\theta_i, \hspace{15pt}\forall i \in \bf{N}.
\end{aligned}\normalsize
\label{equ2}
\end{equation}

\section{Energy Efficiency Optimization Problem}
In this section, we propose and solve an optimization problem to maximize the EE of the WBAN subject to QoS and energy harvesting constraints. Furthermore, the structure of the optimization problem is analyzed to propose a sub-optimal solution at a significantly lower computational complexity. 

\subsection{Problem Formulation and Optimal Solution}
The proposed optimization problem aims to maximize the EE of the WBAN, consisting of SNs equipped with energy harvesting capabilities. The EE objective function is defined as the ratio of the sum of source rate of all the sensors to the power consumption of all the sensors in the network. Mathematically the problem can be formulated as follows
\begin{align}
& \underset{r_i}{\text{maximize}} \quad \frac{ \sum\limits_{i=1}^{n} r_{i}^{(t)}}{\sum\limits_{i=1}^{n} P_{i}^{(t)}}\nonumber \\ 
& \text{subject to}\hspace{13pt} C_1: \hspace{10pt} P_{i}^{(t)}=(\psi_i+X_\sigma\zeta_id^{m_p}_i)r_{i}^{(t)}+\theta_i, \nonumber \hspace{20pt}\forall i,\\ 
&\hspace{49pt}C_2:\quad E_i^{(t+1)}=E^{(t)}_i+\tau\phi_i^{(t)}-\tau P_{i}^{(t)}-F_i^{(t)}, \nonumber\\
&\hspace{49pt}C_3:\quad E^{\min}_i \leq E_i^{(t+1)} \leq E^{\max}_i, \hspace{20pt}\forall i, \nonumber \\
& \hspace{50pt}C_4: \quad r_i^{(t)} \geq 0, \hspace{30pt}\forall i. 
\label{equ3} 
\end{align}
In problem \eqref{equ3}, the constraint $C_1$ represents the total power consumption at the sensor $i$ during time slot $t$. $C_2$ represents the energy at the beginning of the time slot $(t+1)$, and, $E^{(t)}_i$ is the energy of SN $i$ at the beginning of time slot $t$, $\phi_i^{(t)}$ is the energy recharging rate of sensor $i$ at time slot $t$. $F_i^{(t)}$ represents the amount of the energy wasted by the sensor $i$ during time slot $t$ due to battery overflow. $C_3$ shows that the energy at the beginning of time slot $(t+1)$ must not be less than the minimum energy level $E_i^{\min}$ required to be maintained at SN $i$ and should not be larger than the maximum battery capacity $E_i^{\max}$ of sensor $i$.

The optimization problem in \eqref{equ3} is in linear fractional form and can be transformed into an equivalent linear program (LP) with the help of a Charnes-Cooper transformation \cite{hasan2011solving} which is explained as follows. First, we substitute the power of each sensor as given in $C_1$ in the denominator of the objective function. Then, we multiply both the numerator and denominator of the objective function by a positive constant value $\alpha$. That said, the objective function in \eqref{equ3} can be re-written as a standard linear-fractional optimization problem as
\begin{equation}
    \underset{\vec{z}}{\text{maximize}}\quad\frac{\vec{x}^{\,T}\vec{z}}{\vec{e}{\,^T}\vec{z}+f\alpha},
    \label{equ6}
\end{equation}
where $\vec{z}=\vec{r}\alpha$ and $\vec{r}$ is a $n\times1$ vector that contains the source rate of each sensor, i.e., $ \vec{r} = \begin{bmatrix*} r_{1}^{(t)} \hspace{3pt}r_{2}^{(t)} \cdots \hspace{3pt} r_{n}^{(t)} \end{bmatrix*}^T $, and $\vec{x} = \begin{bmatrix*} 1 \hspace{4pt} 1 \hspace{4pt}\cdots \hspace{4pt}1 \end{bmatrix*}^T$. Similarly $\vec{e}$ is a $n\times1$ vector written as follows $ \vec{e} = \begin{bmatrix*} \lambda_{1} \hspace{3pt}\lambda_{2}\hspace{3pt}\cdots \hspace{3pt} \lambda_{n} \end{bmatrix*}^T $, where, $\lambda_i=\psi_i+X_\sigma\zeta_id_i^{m_p}$. The sum of the energy cost of transmit electronics of  the $N$ sensors in the network can be expressed as $f= \theta_1+\theta_2+\hdots+\theta_N$.
The value of $\alpha$ can be selected such that the denominator of the objective function in \eqref{equ6} is equal to one, i.e., ${\vec{e}{\,^T}\vec{z}+f\alpha}=1$, and by substituting the constraints $C_1$ and $C_2$ in $C_{3}$, the equivalent optimization problem of the \eqref{equ3} with linear objective function to maximize the EE of the WBAN can be written as follows
\begin{align}
&\underset{\alpha, \vec{z}}{\text{maximize}} \hspace{5pt} \vec{x}^{\,T} \vec{z} \nonumber\\ 
&\text{subject to} \hspace{9pt} C_1: \hspace{5pt}\alpha E^{\min}_i \leq \alpha E^{(t)}_i-\tau[(\psi_i+X_\sigma\zeta_id^{m_p}_i)z_i+\alpha \theta_i]\nonumber\\&\hspace{65pt}+\alpha \tau\phi^{(t)}_i-\alpha F^{(t)}_i \leq \alpha E^{\max}_i,\nonumber \hspace{20pt}\forall i,\\
& \hspace{45pt}C_2:\hspace{5pt}{\vec{e}{\,^T}\vec{z}+f\alpha}=1, \nonumber\\
&\hspace{45pt}C_3: \hspace{5pt}z_i > 0, \hspace{30pt} \forall i,  \nonumber\\
&\hspace{45pt}C_4:\hspace{5pt}\alpha > 0. 
\label{equ8} 
\end{align}
Furthermore, simplifying the compound inequality constraint $C_1$ in \eqref{equ8}, the ultimate EE problem with the linear objective function and the simplified set of constraints with the decision variables $\vec{z}$ and $\alpha$ can be written in generalized form as follows
\begin{align}
& \underset{\alpha, \vec{z}}{\text{maximize}} \hspace{5pt} \vec{x}^{\,T} \vec{z} \nonumber\\
& \text{subject to} \hspace{8pt}C_1: \hspace{7pt} a_i z_i+ b_i\alpha \leq 0, \nonumber \hspace{15pt} \forall i,\\
& \hspace{44pt}C_2: \hspace{1pt} -a_i z_i+c_i\alpha \leq 0, \nonumber \hspace{15pt} \forall i,\\
& \hspace{44pt}C_3: \hspace{7pt}{\vec{e}{\,^T}\vec{z}+f\alpha}=1, \nonumber\\
&  \hspace{44pt}C_4:\hspace{7pt} z_i>0, \hspace{47pt} \forall i,\nonumber\\ 
&\hspace{44pt}C_5:\hspace{7pt} \alpha > 0.
\label{equ14}
\end{align}
where, $a_i=\tau(\psi_i+X_\sigma\zeta_id^{m_p}_i)$, $b_i=E^{\min}_i+F^{(t)}_i+\tau \theta_i-E^{(t)}_i-\tau\phi^{(t)}_i$, and $c_i$ can be given as follows $c_i=E^{(t)}_i+\tau\phi^{(t)}_i-\tau \theta_i-F^{(t)}_i-E^{\max}_i$.
The optimization problem to maximize the EE of WBAN is now in standard form and can be solved using the simplex method \cite{thie2011introduction}.
\subsection{Sub-optimal Solution}
In this subsection, we exploit the structure of the EE optimization problem for WBAN in \eqref{equ14} and provide a sub-optimal solution with lower computational complexity. The optimization problem finds the source rate of each sensor in the network and based on that information, the power consumption, and ultimately the EE of the overall WBAN is calculated. From the constraints $C_1$ and $C_2$ of the optimization problem defined in \eqref{equ14}, the source rate of the sensor $i$ can be written respectively as
\begin{align}
r_{i}^{(t)} \leq \frac{E^{(t)}_i+\tau\phi^{(t)}_i-\tau \theta_i-F^{(t)}_i-E^{min}_i}{\tau(\psi_i+X_\sigma\zeta_id^{m_p}_i)} = r^{\max}_i,
\label{equ18}
\end{align}
\begin{align}
r_{i}^{(t)} \geq \frac{E^{(t)}_i+\tau\phi^{(t)}_i-\tau \theta_i-F^{(t)}_i-E^{max}_i}{\tau(\psi_i+X_\sigma\zeta_id^{m_p}_i)}=r^{\min}_i.
\label{equ19}
\end{align}
According to \eqref{equ18} and \eqref{equ19}, the source rate of the sensor $i$ can take any value between the minimum $r^{\min}_i$ and the maximum source rate $r^{\max}_i$. By analyzing the optimization problem in \eqref{equ14}, it can be noticed that by relaxing the $C_3$ that is coupling the decision variables in the constraints together, the source rate of each sensor can take either the maximum or minimum source rate values only. In order not to deviate much from the original optimization problem and the optimal solution, the constraint $C_3$ and its effect on the source rates has to be determined. It can be observed that $C_3$ is contributing towards the denominator of our objective function in \eqref{equ6}, where the source rate of each sensor is coupled with the other sensors in the network. 

The main idea of the sub-optimal solution is to choose the source rates such as to maximize the energy efficiency objective function, i.e., to keep the denominator of the objective function as minimum as possible as the numerator is the equal-weighted sum of the sensor source rates. That said, we propose to allocate either the maximum or the minimum source rates to each SN (based on the coefficients $\lambda_i$) to minimize the denominator in \eqref{equ6}. In particular, the SN with higher coefficient $\lambda_i$ should take the minimum source rate and the SN with a lower coefficient $\lambda_i$ will be allocated by the maximum source rate equation. From the system model, the higher $\lambda_i$ means that the sensor is far from the aggregator and on the other hand, the lower coefficient $\lambda_i$ means that the sensor is near to the aggregator. Consequently, it can be concluded from our analysis, that the SN close to the aggregator should transmit its data with maximum source rate $r^{\max}_i$, and the SN far from the aggregator should transmit the data by minimum source rate $r^{\min}_i$.
\begin{algorithm}[ht]
\caption{Proposed sub-optimal algorithm for energy efficiency optimization problem}
\begin{algorithmic}[1]
\STATE \textbf{INPUT} (N, $\bf{R_{min}}, \bf{P_{min}}, \bf{R_{max}}, \bf{P_{max}}$)
\FOR{$i = 0,\dots,n+1$} 
\IF{$i\neq 0$}
\STATE Replace $R_{\max}[i-1]$ with $R_{\min}[i]$
\STATE Replace $P_{\max}[i-1]$ with $P_{\min}[i]$
\ENDIF
\STATE Get sum of all $R_{\min}$ source rates
\STATE Get sum of all $P_{\min}$ power consumption
\STATE Find the energy efficiency
\IF{$i = 0$}
\STATE push energy efficiency value in new array $K[\hspace{2pt} ]$
\ELSE
\STATE push energy efficiency value in new array $K[i]$
\ENDIF
\STATE Increment $i$
\ENDFOR
\STATE \textbf{OUTPUT} Find the maximum energy efficiency from the array $K[i]$ and get the index of the maximum element.
\end{algorithmic}
\label{algo1}
\end{algorithm}
\setlength{\textfloatsep}{10pt}

The question now is to determine the number of SNs transmitting with maximum and minimum rates for a given set of parameters. To address this question, we propose a sub-optimal Algorithm 1 that has twofold objectives: 1) It separates the source rate of each sensor into two groups that either satisfy the maximum or minimum source rate equation. 2) Based on the selection of source rates of each sensor, it finds the maximized EE of the overall WBAN. The proposed algorithm calculates the EE of the WBAN by selecting the optimal source rates combination from either maximum or minimum source rate for each SN, such as to achieve the maximum EE. The basic idea of the algorithm is to assume all the sensors will have a minimum source rate. Then we incrementally assign the maximum source rate for each sensor based on its distance from the aggregator and calculate the EE. The source rate combinations that result in the maximum EE is the required sub-optimal solution. The proposed sub-optimal algorithm is formally summarized at the top of the next page.
\subsection{Complexity Analysis}
The worst case computational complexity of the sub-optimal solution can be analyzed as follows: starting from line 1 of the Algorithm 1, taking input for $N$ number of sensors is independent of any parameters in the optimization problem; therefore its complexity does not scale with the value of $N$. For remaining inputs $\bf{R_{min}}$, $\bf{R_{max}}$, $\bf{P_{min}}$, and $\bf{P_{max}}$ have the complexity of $\mathcal{O}(N)$ each. Since the complexity scales linearly with the number of sensors, the overall complexity of the input is $\mathcal{O}(N)$. The complexity of the \textbf{for} loop at line 2 is $\mathcal{O}(N+2)$ as the loop repeats $N+2$ times. From line 3 to line 6, the overall computational complexity is $\mathcal{O}(N+1)$. For line 7 and line 8, the complexity is $\mathcal{O}(N+1)$ each. From line 9 to line 12, each has a complexity of $\mathcal{O}(1)$ as this computation is independent of $N$. The line 13 executes $N+1$ times, which makes its complexity equals to $\mathcal{O}(N+1)$. Hence, the overall worst-case computational complexity of the sub-optimal energy efficiency algorithm is $\mathcal{O}(N+2)\mathcal{O}(N+1)+\mathcal{O}(N) = \mathcal{O}(N^2)$, which is the polynomial time complexity of $N$. In comparison, the computational complexity of the optimal solution that utilizes the simplex method has a worst-case computational complexity of $\mathcal{O}(2^N)$ \cite{thie2011introduction}.

\section{Results and Discussion}
We considered a WBAN with 10 SNs to evaluate the performance of the proposed optimization problem by performing extensive simulations. By following the simulation parameters of \cite{he2011optimal}, the distance from the SNs to the aggregator has been uniformly distributed between 0.3 and 0.7 m. The initial energy $E^{\mathrm{ini}}$ of each SN is set to 0.1 J. The maximum battery capacity $E^{\mathrm{max}}$ of each sensor is 0.11 J. The minimum energy $E^{\mathrm{min}}$ required for each SN is set as 0.01 J. In the power consumption model, the energy cost of sensing $\psi_i$ and transmit electronics $\theta_i$ for sensor $i$ is set as $2 \times 10^{-8}$ J/b and $6 \times 10^{-8}$ J/b respectively. Similarly, the energy cost of the transmit amplifier $\zeta$ of the sensors is chosen as $8 \times 10^{-8}$ J/b/$\text{m}^{m_p}$. The path loss exponent $m_p$ of the SNs is set between 1.4 to 4.4. In the energy harvesting Markov chain model, from state 1 to state 2, the transition probability is uniformly distributed between 0.6 and 0.8. Similarly, from state 2 to state 1, the transition probability is uniformly distributed between 0.2 and 0.4. The length of the time slot $\tau$ is set as 5 $\text{s}$. In the relaxing state, $X_\sigma$ becomes 1 since during a time slot, the energy recharging rate, the direction, and the distance from a SN to the aggregator remains constant in the relaxing state. However, the arbitrary movements of the human body that change the direction and distance between the SN and the aggregator and cause variations in the path loss are modeled using  $X_\sigma$, denoted as $\mathcal{N}(0, \sigma_s^2)$. Based on the experimental and measured results in \cite{d2010statistical}, the $\sigma_s$ for the walking and the running activities is chosen as 2.15 dB and 3.49 dB, respectively.

\begin{figure*}[t]
\centering
\begin{minipage}[t]{0.323\linewidth} 
\centering
\includegraphics[width=\textwidth]{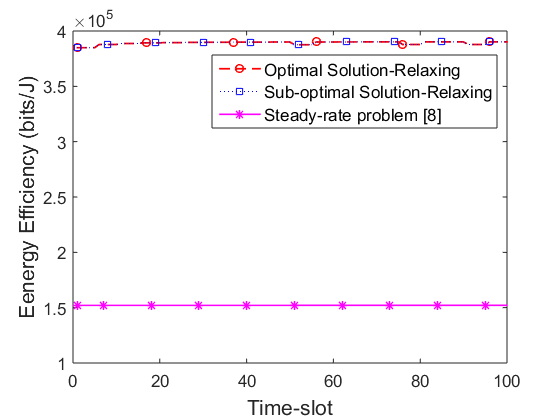}
\caption{\scriptsize Optimal and sub-optimal energy efficiency of WBAN in relaxing state.}
\label{fig1}
\end{minipage}
\hspace{0.04cm}
\begin{minipage}[t]{0.323\linewidth}
\centering
\includegraphics[width=\textwidth]{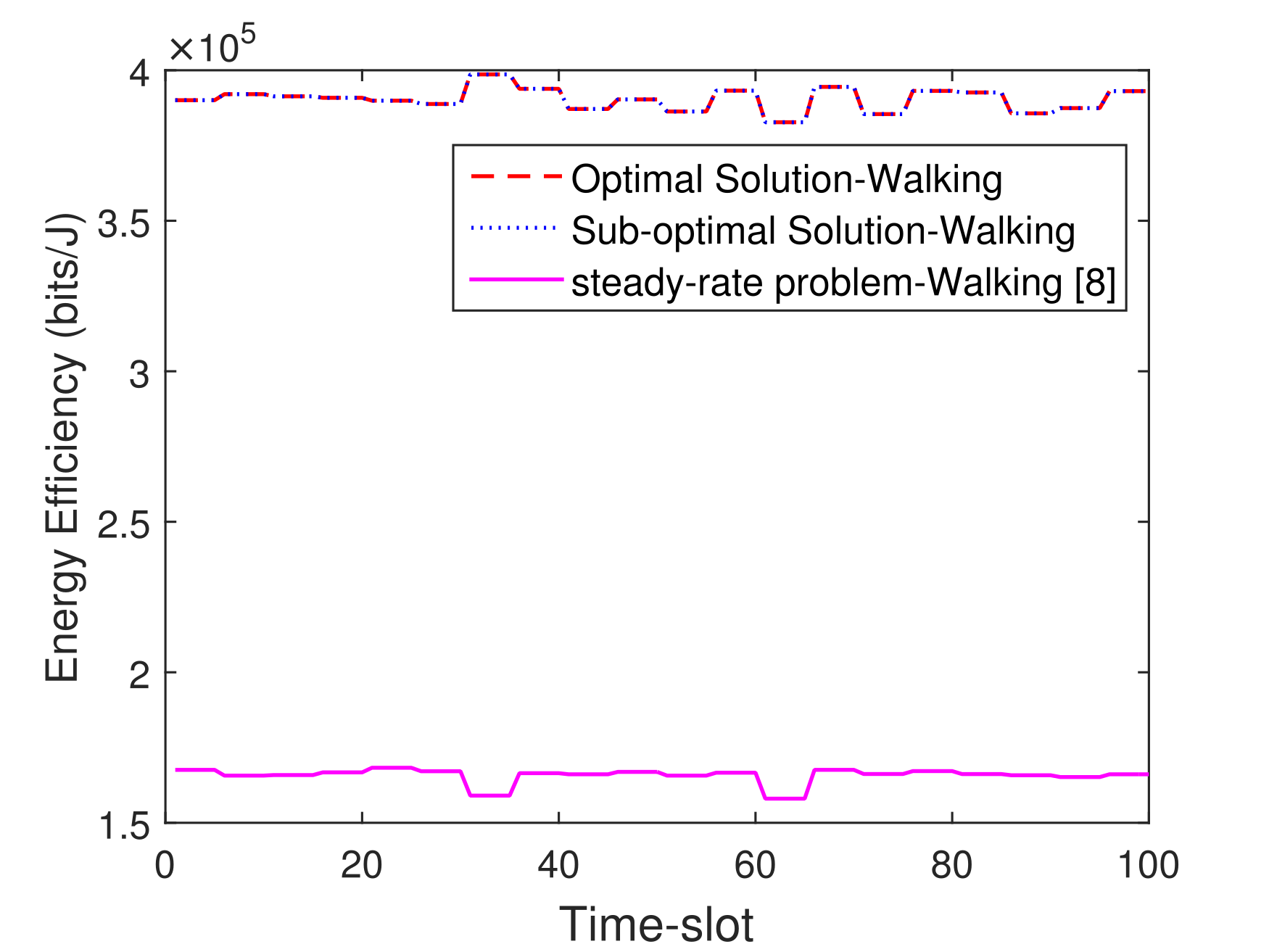}
\caption{\scriptsize Energy efficiency of WBAN in walking state.}
\label{fig2}
\end{minipage}
\hspace{0.04cm}
\begin{minipage}[t]{0.323\linewidth}
\centering
\includegraphics[width=\textwidth]{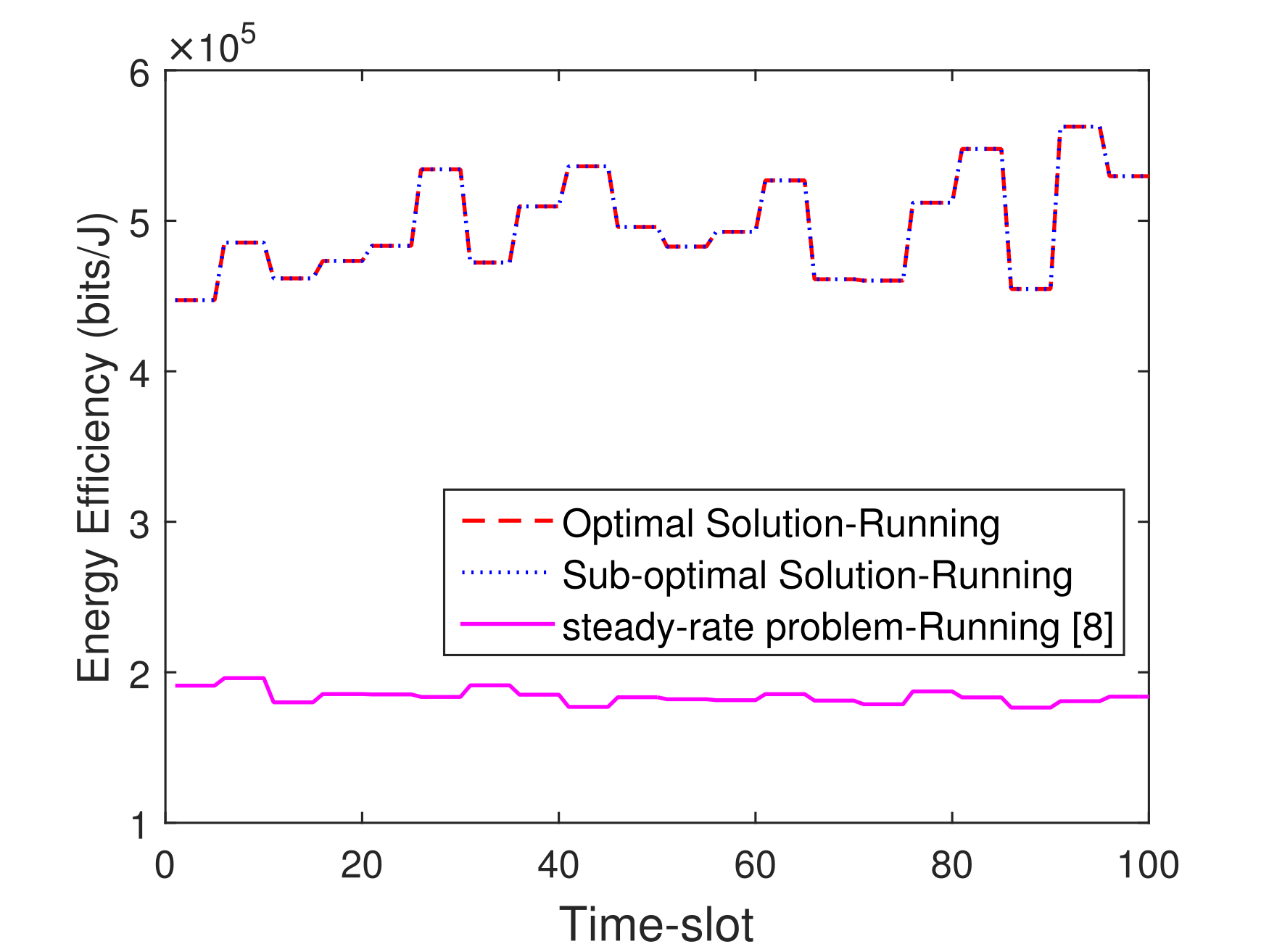}
\caption{\scriptsize Energy efficiency of WBAN in running state.}
\label{fig3}
\end{minipage}
\end{figure*}

Fig.~\ref{fig1} shows the performance of the optimal and sub-optimal solution of the EE optimization problem in relaxing state in comparison with the steady-rate problem \cite{he2011optimal}. It can be noticed that the EE optimization problem provides higher EE in comparison with the steady rate problem. The EE remains almost static during the time slots because the distance and the direction from SN to aggregator remain unchanged in a relaxing state. Moreover, the performance of a sub-optimal solution is very close to the optimal solution.

Fig.~\ref{fig2} illustrates the effect of EE when the patient is in the walking state. The EE varies due to the change of path loss and the distance from the SN to the aggregator. Moreover, the energy harvesting rate also changes due to dynamic behavior. Therefore, for a particular time-slot, if the harvested energy is more than the energy consumed, EE will increase, and if the harvested energy is less than the power consumed, EE will decrease. As depicted from Fig.~\ref{fig3}, in the running state, the EE is higher as compared to the walking state as the harvested energy is higher due to the aggressive body movements.
\section{Conclusion}
In the design of WBAN, due to the limited battery life of the SNs, saving energy is of paramount importance. Therefore, maximizing the EE enables efficient use of the energy critical nodes. This paper formulates a novel optimization problem to maximize the EE of the WBAN with energy harvesting capabilities. The optimization problem is transformed from the linear fractional problem to a linear function, and the resultant problem is solved using numerical methods. For further in-depth analysis, we exploited the structure of the optimization problem and derived the upper and the lower bounds of the source rates, and a sub-optimal solution is proposed that approaches the optimal solution with lower computational complexity. Simulation results validate the proficiency of the proposed schemes, and the performance merits in terms of energy efficiency of the network.

\bibliographystyle{IEEEtran}
\bibliography{IEEEabrv,references}

\begin{thebibliography}{10}

\bibitem{movassaghi2014wireless}
Samaneh Movassaghi, Mehran Abolhasan, Justin Lipman, David Smith, and Abbas
  Jamalipour.
\newblock Wireless body area networks: A survey.
\newblock {\em {IEEE} Commun. Surveys Tuts.}, 16(3):1658--1686, Jan. 2014.

\bibitem{akhtar2017energy}
Fayaz Akhtar and Mubashir~Husain Rehmani.
\newblock Energy harvesting for self-sustainable wireless body area networks.
\newblock {\em {IEEE} {IT} Prof.}, 19(2):32--40, Apr. 2017.

\bibitem{ibarra2016qos}
Ernesto Ibarra, Angelos Antonopoulos, Elli Kartsakli, Joel~JPC Rodrigues, and
  Christos Verikoukis.
\newblock Qo{S}-aware energy management in body sensor nodes powered by human
  energy harvesting.
\newblock {\em {IEEE} Sensors J.}, 16(2):542--549, Jan. 2016.

\bibitem{mekikis2017stochastic}
Prodromos-Vasileios Mekikis, Angelos Antonopoulos, Elli Kartsakli, Nikos
  Passas, Luis Alonso, and Christos Verikoukis.
\newblock Stochastic modeling of wireless charged wearables for reliable health
  monitoring in hospital environments.
\newblock In {\em Proc. {IEEE} International Conference on Communications
  (ICC)}, pages 1--6, May 2017.

\bibitem{yang2018wireless}
Xin Yang, Ling Wang, and Zhaolin Zhang.
\newblock Wireless body area networks {MAC} protocol for energy efficiency and
  extending lifetime.
\newblock {\em IEEE Sensors Lett.}, 2(1):1--4, Jan. 2018.

\bibitem{liu2017medium}
Bin Liu, Zhisheng Yan, and Chang~Wen Chen.
\newblock Medium access control for wireless body area networks with {Q}o{S}
  provisioning and energy efficient design.
\newblock {\em {IEEE} Trans. Mobile Comput.}, 16(2):422--434, Feb. 2017.

\bibitem{esteves2015cooperative}
Volker Esteves, Angelos Antonopoulos, Elli Kartsakli, Manel Puig-Vidal, Pere
  Miribel-Catal{\`a}, and Christos Verikoukis.
\newblock Cooperative energy harvesting-adaptive {MAC} protocol for {WBAN}s.
\newblock {\em Sensors}, 15(6):12635--12650, May 2015.

\bibitem{he2011optimal}
Yifeng He, Wenwu Zhu, and Ling Guan.
\newblock Optimal resource allocation for pervasive health monitoring systems
  with body sensor networks.
\newblock {\em {IEEE} Trans. Mobile Comput.}, 10(11):1558--1575, May 2011.

\bibitem{heinzelman2000energy}
Wendi~Rabiner Heinzelman, Anantha Chandrakasan, and Hari Balakrishnan.
\newblock Energy-efficient communication protocol for wireless microsensor
  networks.
\newblock In {\em Proc. {IEEE} 33rd annual Hawaii international conference on
  system sciences}, pages 1--10, Jan. 2000.

\bibitem{d2010statistical}
Raffaele D'Errico and Laurent Ouvry.
\newblock A statistical model for on-body dynamic channels.
\newblock {\em Int. J. Wirel. Inf. Netw.}, 17(3-4):92--104, Sep. 2010.

\bibitem{hasan2011solving}
Mohammad~Babul Hasan and Sumi Acharjee.
\newblock Solving {LFP} by converting it into a single {LP}.
\newblock {\em Int. J. Ops. Research}, 8(3):1--14, Jun. 2011.

\bibitem{thie2011introduction}
Paul~R Thie and Gerard~E Keough.
\newblock {\em An Introduction to Linear Programming and Game Theory}.
\newblock John Wiley \& Sons, Sep. 2011.

\end{thebibliography}


\end{document}